\documentclass[%
aps,%
showpacs,%
floatfix,%
superscriptaddress,%
nofootinbib,%
preprintnumbers,%
showkeys
]{revtex4}

\usepackage[dvips]{graphicx}
\usepackage{amsmath,amssymb}
\usepackage{color}

\preprint{STUPP-08-198, \\
FTUV-08-1208, \\
IFIC/08-69}



\begin{document}

\title{Model building by coset space dimensional reduction in ten-dimensions
with direct product gauge symmetry}
\date{\today}

\author{Toshifumi Jittoh}
\email{jittoh@krishna.th.phy.saitama-u.ac.jp}
\affiliation{Department of Physics, Saitama University, 
        Shimo-Okubo, Sakura-ku, Saitama, 338-8570, Japan}
\author{Masafumi Koike}
\email{koike@krishna.th.phy.saitama-u.ac.jp}
\affiliation{Department of Physics, Saitama University, 
        Shimo-Okubo, Sakura-ku, Saitama, 338-8570, Japan}
\author{Takaaki Nomura}
\email{nomura@krishna.th.phy.saitama-u.ac.jp}
\affiliation{Department of Physics, Saitama University, 
        Shimo-Okubo, Sakura-ku, Saitama, 338-8570, Japan}
\author{Joe Sato}
\email{joe@phy.saitama-u.ac.jp}
\affiliation{Department of Physics, Saitama University, 
        Shimo-Okubo, Sakura-ku, Saitama, 338-8570, Japan}
\author{Takashi Shimomura}
\email{takashi.shimomura@uv.es}
\affiliation{Departament de F\'{i}sica Te\`{o}rica and IFIC, Universitat de Val\`{e}ncia-CSIC,
E-46100 Burjassot, Val\`{e}ncia, Spain}

\begin{abstract}
We investigate ten-dimensional gauge theories whose extra six-dimensional space is a
compact coset space, $S/R$, and gauge group is a direct product of two Lie groups.
We list up candidates of the gauge group and embeddings of $R$ into them. 
After dimensional reduction of the coset space,
we find fermion and scalar representations
of $G_{\mathrm{GUT}} \times U(1)$ with $G_{\mathrm{GUT}}=SU(5),~SO(10)$ and $E_6$
which accomodate all of the standard model particles.
We also discuss possibilities to generate distinct Yukawa couplings among the generations
using representations with a different dimension for $G_{\mathrm{GUT}}=SO(10)$ and $E_6$ models.
\end{abstract}

\maketitle

\section{Introduction}
\label{sec:Intro}

The Standard Model (SM) has been successful 
in describing phenomenology of the
elementary particle physics up to the energy of order TeV.
Not only did it explain experimental results but 
it also gave us deeper insights
that gauge symmetry governs the interactions among the particles and its
spontaneous breaking rises particle masses.
Despite of its success, the SM is not a satisfactory model because the choice of
the gauge groups and the contents of the particles are the inputs of the model,
and all parameters in Higgs and Yukawa sector,
 which are responsible for the
masses,
 are not predictable.
Grand unification addresses the former points by unifying the gauge symmetries
into single gauge group and fermions into larger representations.
But it requires new scalars to break the grand unification symmetry in the same
manner as the SM,  
resulting in the introduction of more free parameters than those
in the SM.
Therefore a plausible framework for the physics beyond the SM will be an
unification of Higgs and the gauge bosons.

Coset Space Dimensional Reduction (CSDR) scheme is one of the attractive
approaches in this regard~\cite{Manton:1979kb,Forgacs:1979zs,Kapetanakis:1992hf,Chatzistavrakidis:2007by,Zoupanos08,Jittoh:2008jc}.
This scheme introduces a compact extra dimensional space which has the structure
of a coset of Lie groups, $S/R$.
The Higgs field and the gauge field of the SM are merged into a gauge field of a
gauge group $G$ in the higher-dimensional spacetime.
The SM fermions are unified into a representation of this gauge group.
The particle contents surviving in four dimensional theory
are determined by the
identification of the gauge transformation as a rotation within the
extra-dimensional space.
The four dimensional gauge symmetries are determined by embedding of $R$ into
$G$.
Since the Higgs originates from extra dimensional components of the gauge field,
the Higgs and Yukawa sectors in four dimensional Lagrangian are uniquely
determined.
Furthermore, as
 shown in Ref.~\cite{Chapline:1982wy,Chiral-con:K12,Chiral-con:S13,Chiral-con:C14}, it is possible to
obtain chiral fermions when total dimension, $D$, of the spacetime is even.
The chiral fermions can be obtained even from (pseudo)real representations in $D
= 8n + 2$ ($D = 8n + 6$)~\cite{Chapline:1982wy,Chiral-con:C14}.

The case $D = 10$ is the most interesting because the superstring theory, which
is a candidate of a unified theory including gravity, suggests this world exists
in ten-dimensional spacetime.
Thus CSDR models of $D = 10$ can bridge the superstring theory and the SM.
In this spirit, many works have been done in ten dimension, but no realistic
model has emerged yet~\cite{
  Kapetanakis:1992hf,%
  10dim-Model:F2,%
  10dim-Model:D3,%
  10dim-Model:K4,%
  10dim-Model:N5,%
  10dim-Model:K6,%
  10dim-Model:K12,%
  10dim-Model:D14,%
  10dim-Model:B%
}.
A major obstacle to build realistic models is the difficulty to obtain all the
SM fermions.
One of the critical reasons of this difficulty is the smallness of
$\mathrm{SO}(6)$ spinor representation.
Another reason is the small degree of freedom in embedding $R$ into $G$.
These facts strongly restrict the fermion representations surviving in four
dimensions.

In this paper, we introduce a new freedom to the embedding of $R$ into $G$ 
by allowing $G$ to be
a direct product of two Lie groups
in order to overcome the latter difficulty.
%
%
We have more candidates for $G$ and the embeddings of $R$ into them,
%
providing more possibilities to
obtain the SM fermions. 
%
Furthermore, one of the gauge groups can be responsible to the four-dimensional
gauge symmetry while the other can be identified with a family symmetry
\cite{Wu:2008zzj,Appelquist:2006xd,Leontaris:2005gm,Perez:2000fx}, which generates a flavour structure in the Yukawa couplings.
Thus, it is worthwhile to study the CSDR scheme with direct product gauge groups
in ten 
dimensions.
We exhaustively search for fermion contents in the SM and the Grand Unified
Theories (GUTs) with $\mathrm{SU}(5)$, $\mathrm{SO}(10)$ and $E_{6}$, limiting
the dimension of a fermion representation less than 1025.

This paper is organized as follows.
In section 2, we briefly recapitulate the scheme of the coset space
dimensional reduction (CSDR) for the case with a gauge group of ten-dimensional
gauge theory which has direct product structure, and the construction of the
four-dimensional theory by the scheme.
In section 3, we obtain the combinations of the coset space $S/R$ and the
gauge group $G$ of the ten-dimensional theory.
We first obtain the phenomenologically plausible coset space $S/R$ and then we restrict the possible gauge
group $G$ for each $S/R$.
In section 4, we exhaustively list the viable models in four-dimensions.
Section 5 is devoted to 
summary and discussions.

\section{CSDR scheme with direct product gauge group}
\label{sec:Review}

In this section, we briefly recapitulate the scheme of the coset space
dimensional reduction in ten dimensions
with a direct product gauge group~\cite{Kapetanakis:1992hf}.

We begin with a gauge theory defined on a ten-dimensional spacetime $M^{10}$
with a gauge group $G = G_{1} \times G_{2}$ where $G_{1}$ and $G_{2}$ are simple
Lie groups.
Here $M^{10}$ is a direct product of a four-dimensional spacetime $M^4$ and a
compact coset space $S/R$, where $S$ is a compact Lie group and $R$ is a Lie
subgroup of $S$.
The dimension of the coset space $S/R$ is thus $6 \equiv 10 - 4$, implying
$\mathrm{dim} \, S - \dim \, R = 6$.
This structure of extra-dimensional space requires the group $R$ to be embedded
into the group $\mathrm{SO}(6)$, which is a subgroup of the Lorentz group
$\mathrm{SO}(1, 9)$.
Let us denote the coordinates of $M^{10}$ by $X^{M} = (x^{\mu}, y^{\alpha})$,
where $x^{\mu}$ and $y^{\alpha}$ are coordinates of $M^{4}$ and $S/R$,
respectively.
The spacetime index $M$ runs over $\mu \in \{0, 1, 2, 3 \}$ and
$\alpha \in \{4, 5 , \cdots, 9 \}$.
We introduce, in this theory, a gauge field $A_{M}(x, y) = (A_{\mu}(x, y),
A_{\alpha}(x, y))$, which belongs to the adjoint representation of the gauge
group $G$, and fermions $\psi(x, y)$, which lies in a representation $F$ of $G$.

The extra-dimensional space $S/R$ admits $S$ as an isometric transformation
group.
We impose on $A_{M}(X)$ and $\psi(X)$ the following symmetry under this
transformation in order to carry out the dimensional reduction
~\cite{Forgacs:1979zs,Symm-con:W13,Symm-con:J14,Symm-con:O15,Symm-con:Y16,Symm-con:Y16-2}.
Consider a coordinate transformation which acts trivially on $x$ and
gives rise to a $S$-transformation on $y$ as
$(x, y) \rightarrow (x, sy)$, 
where $s \in S$.
We require that the transformation of $A_{M}(X)$ and $\psi(X)$ under this
coordinate transformation should be compensated by a gauge transformation.
This symmetry makes the ten-dimensional Lagrangian invariant under the
$S$-transformation and therefore independent of the coordinate $y$ of $S/R$.
The dimensional reduction is then carried out by integrating over the
coordinate $y$ to obtain the four-dimensional Lagrangian.
The four-dimensional theory consists of the gauge fields $A_{\mu}$,
fermions $\psi$, and in addition the scalar fields originated from $A_{\alpha}$.
The gauge group reduces to a subgroup $H$ of the original gauge
group $G$.

The gauge symmetry and particle contents of the four-dimensional theory are
substantially constrained by the CSDR scheme.
We provide below the prescriptions to identify the four-dimensional gauge group
$H$ and its representations for the particle contents.
First, the gauge group of the four-dimensional theory $H$ is easily identified
as
\begin{equation}
  H = C_{G}(R),
\label{eq:H-is-centralizer}
\end{equation}
where $C_{G}(R)$ denotes the centralizer of $R$ in $G = G_{1} \times G_{2}$~\cite{Forgacs:1979zs}.
Thus the four dimensional gauge group $H$ is determined by the embedding of $R$
into $G$.
We then assume that $R$ has also direct product structure $R = R_{1} \times
R_{2}$ so that we can embed them into $G_{1}$ and $G_{2}$.
Here,
 $R_{1}$ and $R_{2}$ are not 
necessarily simple.
We also assume that four dimensional gauge groups $H$ is obtained from only
$G_{1}$ up to U(1) factors.
This assumption ensures the coupling unification if $H$ is the gauge group of the
SM.
These conditions imply 
\begin{align}
& G = G_1 \times G_2, \\
& R = R_1 \times R_2, \\
& G_1 \supset H \times R_{1}, \\
& G_2 \supset R_{2},
\end{align}
up to U(1) factors.
%
%
%

%
Secondly, the representations of $H$ for the scalar fields are specified by the
following prescription.
Let us decompose the adjoint representation of $S$ according to the embedding $S
\supset R_{1} \times R_{2}$ as,
\begin{align}
\label{dec-S-adj}
\mathrm{adj}~S
&= (\mathrm{adj}~R_1, \mathbf{1}) + (\mathbf{1}, \mathrm{adj}~R_2) + \sum_s ({r_1}_s, {r_2}_s), 
\end{align}
where $r_{1s}$ and $r_{2s}$ are representations of $R_1$ and $R_2$, respectively.
We then decompose the adjoint representation of $G_1$ and $G_2$ according to the
embeddings $G_1 \supset H \times R_1$ and $G_2 \supset R_2$, respectively;
\begin{align}
\label{eq:dec_adjS_direct}
\mathrm{adj}~G_1 &= (\mathrm{adj}~H, \mathbf{1}) + (\mathbf{1}, \mathrm{adj}~R_1) + \sum_{g} (h_g, {r_1}_g) , \\
\mathrm{adj}~G_2 &= \mathrm{adj}~R_2 + \sum_{g} {r_2}_g,
\end{align} 
where $r_{1g}$s and $r_{2g}$s denote representations of $R_1$ and $R_2$, and
$h_{g}$s denote representations of $H$.
The decomposition of $\mathrm{adj} \, G$ thus becomes
\begin{align}
\mathrm{adj}~G 
=& (\mathrm{adj}~G_1, \mathbf{1}) + (\mathbf{1}, \mathrm{adj}~G_2) \nonumber \\
=& (\mathrm{adj}~H, \mathbf{1}, \mathbf{1}) 
	+ (\mathbf{1}, \mathrm{adj}~R_1, \mathbf{1}) 
	+ (\mathbf{1}, \mathbf{1}, \mathrm{adj}~R_2) 
\nonumber \\
&	+ \sum_{g} (h_g, {r_1}_g, \mathbf{1}) 
	+ \sum_{g}(\mathbf{1}, \mathbf{1}, {r_2}_g).
\label{eq:dec_adjG_direct}
\end{align}
The representation of the scalar fields are $h_{g}$s whose
corresponding $(r_{1g},\mathbf{1})$s in the decomposition
Eq.~(\ref{eq:dec_adjG_direct}) are contained also in the set $\{ (r_{1s},r_{2s})\}$ in 
Eq.~(\ref{dec-S-adj}).
Note that the trivial representation $\mathbf{1}$s also remain in
four-dimensions if corresponding $(\mathbf{1},r_{2g})$s of
Eq.~(\ref{eq:dec_adjG_direct}) are also contained in the set $\{(r_{1s},r_{2s})
\}$ in Eq.~(\ref{dec-S-adj}).

Thirdly, the representation of $H$ for the fermion fields 
is
determined as follows~\cite{Manton:1981es}.
Let the group $R$ be embedded into the Lorentz group $\mathrm{SO}(6)$ 
in such a way that the vector representation $\mathbf{6}$ of $\mathrm{SO}(6)$ is
decomposed as 
$\mathbf{6} = \sum_{s} (r_{1s},r_{2s})$,
where $r_{1s}$ and $r_{2s}$ are the representations
obtained in the decomposition Eq.~(\ref{dec-S-adj}).
This embedding specifies a decomposition of the Weyl spinor representations
$\mathbf{4} (\bar{\mathbf{4}})$ of $\mathrm{SO}(6)$ under $\mathrm{SO(6)} \supset R_1 \times R_2$ as
\begin{equation}
  \mathbf{4} = \sum_{i} (\sigma_{1i},\sigma_{2i})
  \biggl(
    \bar{\mathbf{4}} = \sum_{i} (\overline{\sigma}_{1i},\overline{\sigma}_{2i})
  \biggr),
\label{eq:sigmad-decomposition}
\end{equation}
where $\sigma_{1i}(\overline{\sigma}_{1i})$s and $\sigma_{2i}(\overline{\sigma}_{2i})$s are 
irreducible representations of $R_1$ and $R_2$.
We then decompose the $\mathrm{SO}(1,9)$ Weyl spinor $\mathbf{16}$ according to
\begin{math}
  (\mathrm{SU}(2) \times \mathrm{SU}(2)) (\approx \mathrm{SO}(1, 3))
  \times
  \mathrm{SO}(6)
\end{math}
as
\begin{equation}
  \mathbf{16}
  = (\mathbf{2}, \mathbf{1}, \mathbf{4})
  + (\mathbf{1}, \mathbf{2}, \bar{\mathbf{4}}),
\end{equation} 
where $(\mathbf{2}, \mathbf{1})$ and $(\mathbf{1}, \mathbf{2})$ representations
of $\mathrm{SU}(2) \times \mathrm{SU}(2)$ correspond to left- and right-handed
spinors, respectively.
We now decompose a representation $F$ of the gauge group $G$. 
We take $F_{1}$ and $F_{2}$ to be a representation of $G_{1}$ and $G_{2}$ for the
fermions in ten-dimensional spacetime.
Decompositions of $F_{1}$ and $F_{2}$ are
\begin{align}
F_1 &= \sum_f (h_f, {r_1}_f), \\
F_2 &= \sum_f {r_2}_f,
\end{align}
under $G_1 \supset H \times R_1$ and $G_2 \supset R_2$.
Therefore the decomposition of $F$ becomes
\begin{equation}
F = \sum_f (h_f, {r_1}_f, {r_2}_f).
\end{equation}
The representations for the left-handed(right-handed) fermions are $h_{f}$s whose corresponding
$(r_{1f}, r_{2f})$s are found in
\begin{math}
  \{ (\sigma_{1i}, \sigma_{2i}) \}
  ( \{ (\overline{\sigma}_{1i}, \overline{\sigma}_{2i}) \} )
\end{math}
obtained in Eq.~(\ref{eq:sigmad-decomposition}).
Note that a phenomenologically acceptable model needs chiral fermions in the
four dimensions as the SM does.
The chiral fermions are obtained most straightforwardly when we introduce a
complex representation of $G$ as $F$~\cite{Chapline:1982wy,Chiral-con:K12,Chiral-con:S13,Chiral-con:C14}.
More interesting is the possibility to obtain them if $F$ is real
representation, provided $\mathrm{rank} \, S = \mathrm{rank} \, R$~\cite{Bott}.
A pair of Weyl fermions 
appears in a same representation in this case, and one of
the pair is eliminated by imposing the Majorana condition on the Weyl fermions~\cite{Chapline:1982wy,Chiral-con:C14}.
We thus apply the CSDR scheme to complex or real representations of gauge group
$G$ for fermions.
Coset space $S/R$ of our interest 
should satisfy $\mathrm{rank} \, S =
\mathrm{rank} \, R$ to generate chiral fermions in four dimensions~\cite{Bott}.
This condition limits the possible $S/R$ to the coset spaces collected in
Table~\ref{table_6dim_coset_spaces}~\cite{Kapetanakis:1992hf}.
The $R$ of coset (i) in Table~\ref{table_6dim_coset_spaces} with subscript
``max'' indicates that this is the maximal regular subgroup of the $S$.
There,
 the correspondence between the subgroup of $R$ and the
subgroup of $S$ is clarified by the brackets in $R$.
For example, the coset space (iv) suggests direct product of
$\mathrm{Sp}(4) / \mathrm{SU}(2) \times \mathrm{SU}(2)$ and
$\mathrm{SU}(2) / \mathrm{U}(1)$.

Here we mention the effect of gravity.
When we include the effect of gravity and consider 
dynamics of an extra-space we would find the difficulty to obtain stable extra space.
This is the common difficulty of extra-dimensional models 
and some works  have been done on this point.
For example it is discussed in terms of radion fields 
which are the scalar fields originated from higher-dimensional components of metric 
after compactification [31],[32].
The effect of gravity to CSDR scheme is also discussed in [4], [5].
Although we agree that the effect of gravity is important,
we do not discuss about the effect of gravity in this letter since
it is beyond the scope of this letter.

\begin{table}
  \caption{%
    A complete list of six-dimensional coset spaces $S/R$ with
    $\mathrm{rank} \, S = \mathrm{rank} \, R$~\cite{Kapetanakis:1992hf}.
    The brackets in $R$ clarifies the
    correspondence between the subgroup of $R$ and the subgroup of $S$.
    The factor of $R$ with subscript ``max'' indicates that this factor is
    a maximal regular subgroup of $S$.
    }
    \label{table_6dim_coset_spaces}
\begin{center}
\begin{scriptsize}
\begin{tabular}{ll}
  \hline
  No. & $S/R$
  \\ \hline \hline
  (i)
  & $\mathrm{Sp}(4) / [\mathrm{SU}(2) \times \mathrm{U}(1)]_\mathrm{max}$
  \\ \hline
  (ii)
  & $\mathrm{Sp}(4) / [\mathrm{SU}(2) \times \mathrm{U}(1)]_\textrm{non-max}$
  \\ \hline
  (iii)
  & $\mathrm{SU}(4) / \mathrm{SU}(3) \times \mathrm{U}(1)$
  \\ \hline
  (iv)
  & $\mathrm{Sp}(4) \times \mathrm{SU}(2) / [\mathrm{SU}(2) \times \mathrm{SU}(2)] \times \mathrm{U}(1)$
  \\ \hline
  (v)
  & $\mathrm{G}(2) / \mathrm{SU}(3)$
  \\ \hline
  (vi)
  & $\mathrm{SO}(7) / \mathrm{SO}(6)$
  \\ \hline
  (vii)
  & $\mathrm{SU}(3) / \mathrm{U}(1) \times \mathrm{U}(1)$
  \\ \hline
  (viii)
  & $\mathrm{SU}(3) \times \mathrm{SU}(2) / [\mathrm{SU}(2) \times \mathrm{U}(1)] \times \mathrm{U}(1)$
  \\ \hline
  (ix)
  & $\left(\mathrm{SU}(2) / \mathrm{U}(1) \right)^3$
  \\ \hline
\end{tabular}
\end{scriptsize}
\end{center}
\end{table}

%
%
\section{Candidates of the coset space $S/R$ and the gauge group $G$}
In this section we obtain the combinations of the coset space $S/R$ and the
gauge group $G$ of the ten-dimensional theory.
We first obtain the coset space $S/R$ and then we restrict the possible gauge
group $G$ for each $S/R$.

We select the coset space $S/R$ from the ones listed in
Table~\ref{table_6dim_coset_spaces} by the following two criteria.
First, $R$ should be a direct product of subgroups $R_{1}$ and $R_{2}$ to have
new freedom to embedding of $R$ into $G$.
This criterion exculdes the candidates of $S/R$ (v) and (vi) in
Table~\ref{table_6dim_coset_spaces}.

Secondly, the four-dimensional gauge group obtained by
Eq.~(\ref{eq:H-is-centralizer}) should be that of the SM or a GUT with at most
one extra $\mathrm{U}(1)$ gauge group, \textit{i.e.} the SM-like gauge group
$G_{\mathrm{SM}} (\times \mathrm{U}(1))$, where
\begin{math}
  G_{\mathrm{SM}} \equiv
  \mathrm{SU}(3) \times \mathrm{SU}(2) \times \mathrm{U}(1),
\end{math}
or a GUT-like gauge group $G_{\textrm{GUT}} (\times \mathrm{U}(1))$, where
$G_{\textrm{GUT}}$ is either $\mathrm{SU}(5)$, $\mathrm{SO}(10)$ or $\mathrm{E}_{6}$.
This criterion exculdes the candidates (vii) -- (ix) in
Table~\ref{table_6dim_coset_spaces} by the following reasons.
%
%
%
%
\begin{enumerate}
\item We note that the $\mathrm{U}(1)$s in $R$ are also parts of its
  centralizer, \textit{i.e.} a part of $H$.
  We thus exclude the candidate (ix) since we consider the $H$s which have at
  most two $\mathrm{U}(1)$ factors.
\item Similarly, as long as we consider the GUT-like and $G_{\textrm{SM}}$ gauge groups, we do not need
  to consider the candidates (vii) and (viii).
\item The candidates (vii) and (viii) 
do not allow $H = G_{\textrm{SM}} \times
  \mathrm{U}(1)$ either for the following reason.
  The hypercharge of the SM should be reproduced by a certain linear combination
  of two $\mathrm{U}(1)$s in $R$, which should be matched to the
  spinor representation of $\mathrm{SO}(6)$.
  The dimension of the $\mathrm{SO}(6)$ spinor representation is four, and thus
  no more than four different values of $\mathrm{U}(1)$ charges are available.
  On the other hand the fermion content of the SM has five different values of
  $\mathrm{U}(1)$ charges.
  Hence,
 this case never reproduces the hypercharges of the SM fermions.
  %
\item Due to the above three reasons 
the candidates (i) -- (iv) allow neither $G_{\textrm{SM}}$ nor $G_{\textrm{SM}}$ as $H$.
\end{enumerate}  
To summarize, the possible model requires coset space $S/R$ listed in (i) -- (iv)
of Table~\ref{table_6dim_coset_spaces}, with either $H = G_{\mathrm{SM}} \times
\mathrm{U}(1)$ or $H = G_{\mathrm{GUT}} \times \mathrm{U}(1)$.
In Table~\ref{so6_vector_and_spinor} we show the embedding of $R$ in
$\mathrm{SO}(6)$ for these coset spaces.
The representations of $r_{s}$ in Eq.~(\ref{dec-S-adj}) and $\sigma_{i}$ in
Eq.~(\ref{eq:sigmad-decomposition}) are listed in the columns of
``Branches of $\mathbf{6}$'' and ``Branches of $\mathbf{4}$'',
respectively. 
\begin{table}
  \caption{%
    The decompositions of the vector representation $\mathbf{6}$
    and the spinor representation $\mathbf{4}$ of $\mathrm{SO}(6)$
    under $R$s which are listed as (i) --(iv) in
    Table~\ref{table_6dim_coset_spaces}.
    The representations of $r_{s}$ in Eq.~(\ref{dec-S-adj}) and $\sigma_{i}$ 
    in Eq.~(\ref{eq:sigmad-decomposition}) 
    are listed in the columns of ``Branches of $\mathbf{6}$'' and 
    ``Branches of $\mathbf{4}$'', respectively. 
    %
  }
  \label{so6_vector_and_spinor}
\begin{scriptsize}
\begin{center}
  \renewcommand{\arraystretch}{1.2}
\begin{tabular}{llll}
  \hline
  $S/R$
  &
  & Branches of $\mathbf{6}$
  & Branches of $\mathbf{4}$
  \\ \hline \hline
  (i) 
  & $\mathrm{SU}(2)(\mathrm{U}(1))$
  & $\mathbf{3}(2), \mathbf{3}(-2)$ 
  & $\mathbf{1}(3), \mathbf{3}(-1)$ 
  \\
  \hline
   (ii) 
  & $\mathrm{SU}(2)(\mathrm{U}(1))$
  & $\mathbf{1}(2), \mathbf{1}(-2), \mathbf{2}(1), \mathbf{2}(-1)$ 
  & $\mathbf{2}(1), \mathbf{1}(0), \mathbf{1}(-2)$ 
  \\
  \hline
  (iii)
  & $\mathrm{SU}(3)(\mathrm{U}(1))$
  & $\mathbf{3}(-4), \bar{\mathbf{3}}(4)$ 
  & $\mathbf{1}(-6), \mathbf{3}(2)$ 
  \\
  \hline
  (iv)
  & $(\mathrm{SU}(2), \mathrm{SU}(2))(\mathrm{U}(1))$
  & $(\mathbf{2}, \mathbf{2})(0), (\mathbf{1}, \mathbf{1})(2), (\mathbf{1}, \mathbf{1})(-2)$ 
  & $(\mathbf{2}, \mathbf{1})(1), (\mathbf{1}, \mathbf{2})(-1)$ 
  \\ 
  \hline
\end{tabular}
\end{center}
\end{scriptsize}
\end{table}
%
%
%
%
%
%
The embedding of $R$ into higher dimensional gauge group $G = G_{1} \times
G_{2}$ 
is listed in Table~\ref{R_embeding_12}--\ref{R_embeding_4}.
These embeddings are straightforwardly obtained by decomposing gauge group $G$ to 
its regular subgroup which contains an $R$-subgroup of $G$. 
A detailed discussion about the embeddings is summarized in \cite{Jittoh:2008jc}.
For each embedding of $R$, the candidates of $G$ are summarized in
Table~\ref{G_candidates_12}--\ref{G_candidates_4-cd}.
Note that all the candidates of $G$ in Table~\ref{G_candidates_12}--\ref{G_candidates_4-cd} 
are subgroup of SO$(32)$ or $\mathrm{E}_8 \times \mathrm{E}_8$ which are required by superstring theory.
%
%
%
\begin{table}
  \caption{
    The embedding of $R$ into $G = G_{1} \times G_2$ for the coset spaces
    (i) and (ii).  
  }
 \label{R_embeding_12}
\begin{center}
\begin{scriptsize}
\begin{tabular}{rll}
\hline
  \multicolumn{3}{l}{
  (i) $\mathrm{Sp}(4) / [\mathrm{SU}(2) \times \mathrm{U}(1)]_\mathrm{max}$ and 
  (ii) $\mathrm{Sp}(4) / [\mathrm{SU}(2) \times \mathrm{U}(1)]_\textrm{non-max}$ 
  }
  \\ \hline \hline
  (a)
  &$G_1 \supset (G_{\mathrm{SM}}~\mathrm{or}~G_{\mathrm{GUT}}) \times \mathrm{SU}(2)$,
  &$G_2 \supset \mathrm{U}(1)$
  \\ \hline 
  (b)
  &$G_1 \supset (G_{\mathrm{SM}}~\mathrm{or}~G_{\mathrm{GUT}}) \times \mathrm{U}(1)$,
  &$G_2 \supset \mathrm{SU}(2)$ 
  \\ \hline
\end{tabular}
\end{scriptsize}
\end{center}
\end{table}
\begin{table}
  \caption{
    The embedding of $R$ into $G = G_{1} \times G_2$ for the coset space
    (iii). 
  }
 \label{R_embeding_3}
\begin{center}
\begin{scriptsize}
\begin{tabular}{rll}
\hline
  \multicolumn{3}{l}{
  (iii) $\mathrm{SU}(4) / \mathrm{SU}(3) \times \mathrm{U}(1)$
  }
  \\ \hline \hline
  (a)
  &$G_1 \supset (G_{\mathrm{SM}}~\mathrm{or}~G_{\mathrm{GUT}}) \times \mathrm{SU}(3)$,
  &$G_2 \supset \mathrm{U}(1)$
  \\ \hline 
  (b)
  &$G_1 \supset (G_{\mathrm{SM}}~\mathrm{or}~G_{\mathrm{GUT}}) \times \mathrm{U}(1)$,
  &$G_2 \supset \mathrm{SU}(3)$ 
  \\ \hline
\end{tabular}
\end{scriptsize}
\end{center}
\end{table}
\begin{table}
  \caption{
    The embedding of $R$ into $G = G_{1} \times G_2$ for the coset space (iv).  
  }
 \label{R_embeding_4}
\begin{center}
\begin{scriptsize}
\begin{tabular}{rll}
\hline
  \multicolumn{3}{l}{
  (iv) $\mathrm{Sp}(4) \times \mathrm{SU}(2) / [\mathrm{SU}(2) \times \mathrm{SU}(2)] \times \mathrm{U}(1)$
  }
  \\ \hline \hline
  (a)
  &$G_1 \supset (G_{\mathrm{SM}}~\mathrm{or}~G_{\mathrm{GUT}}) \times \mathrm{SU}(2)$,
  &$G_2 \supset \mathrm{SU}(2) \times \mathrm{U}(1)$
  \\ \hline 
  (b)
  &$G_1 \supset (G_{\mathrm{SM}}~\mathrm{or}~G_{\mathrm{GUT}}) \times \mathrm{SU}(2) \times \mathrm{SU}(2)$,
  &$G_2 \supset \mathrm{U}(1)$
  \\ \hline 
  (c)
  &$G_1 \supset (G_{\mathrm{SM}}~\mathrm{or}~G_{\mathrm{GUT}}) \times \mathrm{U}(1)$,
  &$G_2 \supset \mathrm{SU}(2) \times \mathrm{SU}(2)$ 
  \\ \hline
  (d)
  &$G_1 \supset (G_{\mathrm{SM}}~\mathrm{or}~G_{\mathrm{GUT}}) \times \mathrm{SU}(2) \times \mathrm{U}(1)$,
  &$G_2 \supset \mathrm{SU}(2)$
  \\ \hline 
\end{tabular}
\end{scriptsize}
\end{center}
\end{table}
\begin{table}
  \caption{
    The candidates of the gauge groups $G_1$ and $G_2$ for each of
    the coset space (i) and (ii) 
    in Table~\ref{table_6dim_coset_spaces}.
    The top row indicates the assigned number of $S/R$ in
    Table~\ref{table_6dim_coset_spaces} 
    and embedding of $R$ assigned in Table~\ref{R_embeding_12}.  
    The leftmost column indicates $H$.
  }
 \label{G_candidates_12}
\begin{center}
\begin{scriptsize}
\begin{tabular}{l|l|l}
  \hline
  & (i)-(a) and (ii)-(a)
  & (i)-(b) and (ii)-(b)
  \\ \hline \hline
  $\mathrm{SU}(3) \times \mathrm{SU}(2) \times \mathrm{U}(1) \times \mathrm{U}(1)$
  & $G_1=$ $\mathrm{SO}(10)$, $\mathrm{SO}(11)$,
  & $G_1=$ $\mathrm{SU}(6)$, $\mathrm{SO}(10)$, 
  \\[-3pt]
  & \hspace{7mm} $\mathrm{Sp}(10)$
  & \hspace{7mm} $\mathrm{SO}(11)$, $\mathrm{Sp}(10)$
  \\
  & $G_2=$ $\mathrm{SU}(2)$, $\mathrm{U}(1)$
  & $G_2=$ $\mathrm{SU}(2)$
  \\ \hline  
  $\mathrm{SU}(5) \times \mathrm{U}(1)$
  & $G_1=$ No candidate
  & $G_1=$ $\mathrm{SU}(6)$, $\mathrm{SO}(10)$, 
  \\[-3pt]
  & 
  & \hspace{7mm} $\mathrm{SO}(11)$, $\mathrm{Sp}(10)$
  \\
  & $G_2=$ $\mathrm{SU}(2)$, $\mathrm{U}(1)$
  & $G_2=$ $\mathrm{SU}(2)$
  \\ \hline 
  $\mathrm{SO}(10) \times \mathrm{U}(1)$
  & $G_1=$ $\mathrm{SO}(13)$
  & $G_1=$ $\mathrm{SO}(12)$, $\mathrm{SO}(13)$, 
  \\[-3pt]
  & 
  & \hspace{7mm} $\mathrm{E}_6$
  \\
  & $G_2=$ $\mathrm{SU}(2)$, $\mathrm{U}(1)$
  & $G_2=$ $\mathrm{SU}(2)$
  \\ \hline
  $\mathrm{E}_{6} \times \mathrm{U}(1)$
  & $G_1=$ No candidate
  & $G_1=$ $\mathrm{E}_7$
  \\
  & $G_2=$ $\mathrm{SU}(2)$, $\mathrm{U}(1)$
  & $G_2=$ $\mathrm{SU}(2)$
  \\ \hline
\end{tabular}
\end{scriptsize}
\end{center}
\end{table}
\begin{table}
  \caption{
    The allowed candidates of the gauge groups $G_1$ and $G_2$ for the
    coset space (iii) 
    in Table~\ref{table_6dim_coset_spaces}.
    The top row indicates the assigned number of $S/R$ in Table~\ref{table_6dim_coset_spaces} 
    and embedding of $R$ assigned in Table~\ref{R_embeding_3}.  
    The leftmost column indicates $H$.
  }
 \label{G_candidates_3}
\begin{center}
\begin{scriptsize}
\begin{tabular}{l|l|l}
  \hline
  & (iii)-(a)
  & (iii)-(b)
  \\ \hline \hline
  $\mathrm{SU}(3) \times \mathrm{SU}(2) \times \mathrm{U}(1) \times \mathrm{U}(1)$
  &$G_1=$ $\mathrm{E}_6$
  & $G_1=$ $\mathrm{SU}(6)$, $\mathrm{SO}(10)$,
  \\[-3pt]
  & 
  & \hspace{7mm} $\mathrm{SO}(11)$, $\mathrm{Sp}(10)$
  \\
  & $G_2=$ $\mathrm{SU}(2)$, $\mathrm{U}(1)$
  & $G_2=$ $\mathrm{G}_2$, $\mathrm{SU}(3)$
  \\ \hline  
  $\mathrm{SU}(5) \times \mathrm{U}(1)$
  & $G_1=$ No candidate
  & $G_1=$ $\mathrm{SU}(6)$, $\mathrm{SO}(10)$,
  \\[-3pt]
  &
  & \hspace{7mm} $\mathrm{SO}(11)$, $\mathrm{Sp}(10)$
  \\
  & $G_2=$ $\mathrm{SU}(2)$, $\mathrm{U}(1)$
  & $G_2=$ $\mathrm{G}_2$, $\mathrm{SU}(3)$
  \\ \hline 
  $\mathrm{SO}(10) \times \mathrm{U}(1)$
  & $G_1=$  No candidate
  & $G_1=$ $\mathrm{SO}(12)$, $\mathrm{SO}(13)$
  \\[-3pt]
  & 
  & \hspace{7mm} $\mathrm{E}_6$
  \\
  & $G_2=$ $\mathrm{SU}(2)$, $\mathrm{U}(1)$
  & $G_2=$ $\mathrm{G}_2$, $\mathrm{SU}(3)$
  \\ \hline
  $\mathrm{E}_{6} \times \mathrm{U}(1)$
  & $G_1=$ $\mathrm{E}_8$
  & $G_1=$ $\mathrm{E}_7$
  \\
  & $G_2=$ $\mathrm{SU}(2)$, $\mathrm{U}(1)$
  & $G_2=$ $\mathrm{G}_2$, $\mathrm{SU}(3)$
  \\ \hline
\end{tabular}
\end{scriptsize}
\end{center}
\end{table}
\begin{table}
  \caption{
    The allowed candidates of the gauge groups $G_1$ and $G_2$ for the
    coset space (iv) 
    in Table~\ref{table_6dim_coset_spaces}.
    The top row indicates the assigned number of $S/R$ in Table~\ref{table_6dim_coset_spaces} 
    and embedding of $R$ assigned in Table~\ref{R_embeding_4}.  
    The leftmost column indicates $H$.
  }
 \label{G_candidates_4-ab}
\begin{center}
\begin{scriptsize}
\begin{tabular}{l|l|l}
  \hline
  & (iv)-(a)
  & (iv)-(b)
  \\ \hline \hline
  $\mathrm{SU}(3) \times \mathrm{SU}(2) \times \mathrm{U}(1) \times \mathrm{U}(1)$
  & $G_1=$ $\mathrm{SO}(10)$, $\mathrm{SO}(11)$, 
  & $G_1=$ $\mathrm{SO}(13)$, $\mathrm{Sp}(12)$
  \\[-3pt]
  & \hspace{7mm} $\mathrm{Sp}(10)$
  & 
  \\
  & $G_2=$ $\mathrm{SU}(3)$, $\mathrm{Sp}(4)$,
  & $G_2=$ $\mathrm{SU}(2)$, $\mathrm{U}(1)$
  \\[-3pt]
  & \hspace{7mm} $\mathrm{G}_2$
  & 
  \\ \hline           
  $\mathrm{SU}(5) \times \mathrm{U}(1)$
  & $G_1=$ No candidate
  & $G_1=$ No candidate
  \\
  & $G_2=$ $\mathrm{SU}(3)$, $\mathrm{Sp}(4)$,
  & $G_2=$ $\mathrm{SU}(2)$, $\mathrm{U}(1)$
  \\[-3pt]
  & \hspace{7mm} $\mathrm{G}_2$
  &
  \\ \hline 
  $\mathrm{SO}(10) \times \mathrm{U}(1)$
  & $G_1=$ $\mathrm{SO}(13)$
  & $G_1=$ $\mathrm{SO}(14)$, $\mathrm{SO}(15)$
  \\
  & $G_2=$ $\mathrm{SU}(3)$, $\mathrm{Sp}(4)$,
  & $G_2=$ $\mathrm{SU}(2)$, $\mathrm{U}(1)$
  \\[-3pt]
  & \hspace{7mm} $\mathrm{G}_2$
  &
  \\ \hline
  $\mathrm{E}_{6} \times \mathrm{U}(1)$
  & $G_1=$ No candidate
  & $G_1=$ No candidate
  \\
  & $G_2=$ $\mathrm{SU}(3)$, $\mathrm{Sp}(4)$, 
  & $G_2=$ $\mathrm{SU}(2)$, $\mathrm{U}(1)$
  \\[-3pt]
  & \hspace{7mm} $\mathrm{G}_2$
  &
  \\ \hline
\end{tabular}
\end{scriptsize}
\end{center}
\end{table}
\begin{table}
  \caption{
    The allowed candidates of the gauge groups $G_1$ and $G_2$ for the
    coset space (iv) 
    in Table~\ref{table_6dim_coset_spaces}.
    The top row indicates the assigned number of $S/R$ in Table~\ref{table_6dim_coset_spaces} 
    and embedding of $R$ assigned in Table~\ref{R_embeding_4}.  
    The leftmost column indicates $H$.
  }
 \label{G_candidates_4-cd}
\begin{center}
\begin{scriptsize}
\begin{tabular}{l|l|l}
  \hline
  & (iv)-(c)
  & (iv)-(d)
  \\ \hline \hline
  $\mathrm{SU}(3) \times \mathrm{SU}(2) \times \mathrm{U}(1) \times \mathrm{U}(1)$
  & $G_1=$ $\mathrm{SU}(6)$, $\mathrm{SO}(10)$,
  & $G_1=$ $\mathrm{SU}(7)$, $\mathrm{SO}(12)$,
  \\[-3pt]
  & \hspace{7mm} $\mathrm{SO}(11)$, $\mathrm{Sp}(10)$
  & \hspace{7mm} $\mathrm{SO}(13)$, $\mathrm{Sp}(12)$,
  \\[-3pt]
  &
  & \hspace{7mm} $\mathrm{E}_6$
  \\
  & $G_2=$ $\mathrm{G}_2$, $\mathrm{Sp}(4)$
  & $G_2=$ $\mathrm{SU}(2)$
  \\ \hline  
  $\mathrm{SU}(5) \times \mathrm{U}(1)$
  & $G_1=$ $\mathrm{SU}(6)$, $\mathrm{SO}(10)$,
  & $G_1=$ $\mathrm{SU}(7)$, $\mathrm{SO}(13)$
  \\[-3pt]
  & \hspace{7mm} $\mathrm{SO}(11)$, $\mathrm{Sp}(10)$
  & \hspace{7mm} $\mathrm{Sp}(12)$, $\mathrm{E}_6$
  \\
  & $G_2=$ $\mathrm{G}_2$, $\mathrm{Sp}(4)$
  & $G_2=$ $\mathrm{SU}(2)$
  \\ \hline 
  $\mathrm{SO}(10) \times \mathrm{U}(1)$
  & $G_1=$ $\mathrm{SO}(12)$, $\mathrm{SO}(13)$,
  & $G_1=$ $\mathrm{SO}(14)$, $\mathrm{SO}(15)$,
  \\[-3pt]
  & \hspace{7mm} $\mathrm{E}_6$
  & \hspace{7mm} $\mathrm{E}_7$
  \\
  & $G_2=$ $\mathrm{G}_2$, $\mathrm{Sp}(4)$
  & $G_2=$ $\mathrm{SU}(2)$
  \\ \hline
  $\mathrm{E}_{6} \times \mathrm{U}(1)$
  & $G_1=$ $\mathrm{E}_7$
  & $G_1=$ $\mathrm{E}_8$
  \\
  & $G_2=$ $\mathrm{G}_2$, $\mathrm{Sp}(4)$
  & $G_2=$ $\mathrm{SU}(2)$
  \\ \hline
\end{tabular}
\end{scriptsize}
\end{center}
\end{table}
%

%
%

The representation $F_{1}$ of $G_{1}$ for the fermions should be either complex
or real but not pseudoreal, since the fermions of pseudoreal representation do
not allow the Majorana condition when $D = 10$ and 
induce doubled fermion
contents after the dimensional reduction~\cite{Chapline:1982wy}.
Table~\ref{representations} lists the candidate groups $G_{1}$ and their complex
and real representations whose dimension is no more than 1024.
The representations in this table are the candidates of $F_{1}$.
The groups $\mathrm{SU}(7)$ and $\mathrm{SO}(13)$ are not listed
here since they do not lead to the four-dimensional gauge
group of our interest for any of $S/R$ and embedding of $R$ in
Table~\ref{R_embeding_12}--\ref{R_embeding_4}.

The representation $F_{2}$ of $G_{2}$ has to be real as well as $F_{1}$ to
impose the Majorana condition.
Without this condition, $F_{2}$ can be any representation.
%
%
We limited ourselves to the case
\begin{math}
  \mathrm{dim} \, F = \mathrm{dim} \, F_1 \times \mathrm{dim} \, F_2 < 1025
\end{math}
since larger representations yield numerous higher dimensional representations
of fermion under the $G_{\textrm{SM}} \times U(1)$ and $G_{\textrm{GUT}} \times U(1)$.

\begin{table}
  \caption{
    The complex or real
    representations of the possible gauge groups
    \cite{MckayPatera}.
    The groups $\mathrm{SU}(7)$ and $\mathrm{SO}(13)$ are not listed
    here since they do not lead to the four-dimensional gauge
	group of our interest for any of $S/R$ and embedding of $R$ in
	Table~\ref{R_embeding_12}--\ref{R_embeding_4}.
  }
\label{representations}
\begin{center}
\begin{scriptsize}
\begin{tabular}{lll}
  \hline
  Group
  & Complex representations
  & Real representations
  \\ \hline \hline
$\mathrm{SU}(6)$
  & 
  $\mathbf{6}$, 
  $\mathbf{15}$, 
  $\mathbf{21}$,
  $\mathbf{56}$, 
  $\mathbf{70}$, 
  $\mathbf{84}$, 
  $\mathbf{105}$, 
  $\mathbf{105'}$, 
  $\mathbf{120}$, 
  & 
  $\mathbf{35}$, 
  $\mathbf{175}$, 
  $\mathbf{189}$, 
  $\mathbf{405}$, 
  $\cdots$ 
  \\
  & 
  $\mathbf{126}$, 
  $\mathbf{210}$, 
  $\mathbf{210'}$, 
  $\mathbf{252}$, 
  $\mathbf{280}$, 
  $\mathbf{315}$, 
  $\mathbf{336}$, 
  $\mathbf{384}$, 
  & 
  \\
  &  
  $\mathbf{420}$, 
  $\mathbf{462}$, 
  $\mathbf{490}$,
  $\mathbf{504}$, 
  $\mathbf{560}$, 
  $\mathbf{700}$, 
  $\mathbf{720}$, 
  $\mathbf{792}$, 
  &
  \\
  & 
  $\mathbf{840}$, 
  $\mathbf{840'}$,
  $\mathbf{840''}$, 
  $\mathbf{896}$, 
  $\cdots$ 
  & 
  \\ \hline
  %
  %
$\mathrm{SO}(11)$
  & 
  & 
  $\mathbf{11}$, 
  $\mathbf{55}$, 
  $\mathbf{65}$, 
  $\mathbf{165}$, 
  $\mathbf{275}$, 
  $\mathbf{320}$, 
  $\mathbf{330}$, 
  $\mathbf{429}$, 
  \\
  & 
  & 
  $\mathbf{462}$, 
  $\mathbf{935}$, 
  $\cdots$ 
  \\ \hline
$\mathrm{SO}(12)$
  & 
  & 
  $\mathbf{12}$, 
  $\mathbf{66}$, 
  $\mathbf{77}$, 
  $\mathbf{220}$, 
  $\mathbf{352}$, 
  $\mathbf{462}$, 
  $\mathbf{495}$, 
  $\mathbf{560}$, 
  \\
  & 
  & 
  $\mathbf{792}$, 
  $\cdots$ 
  \\ \hline
  %
  %
$\mathrm{SO}(14)$
  & 
  $\mathbf{64}$, 
  $\mathbf{832}$, 
  $\cdots$ 
  & 
  $\mathbf{14}$, 
  $\mathbf{91}$, 
  $\mathbf{104}$, 
  $\mathbf{364}$, 
  $\mathbf{546}$, 
  $\mathbf{896}$, 
  $\cdots$ 
  \\ \hline
$\mathrm{SO}(15)$
  & 
  & 
  $\mathbf{15}$, 
  $\mathbf{105}$, 
  $\mathbf{119}$, 
  $\mathbf{128}$, 
  $\mathbf{455}$, 
  $\mathbf{665}$, 
  $\cdots$ 
  \\ \hline
$\mathrm{F_4}$
  & 
  & 
  $\mathbf{26}$, 
  $\mathbf{52}$, 
  $\mathbf{273}$, 
  $\mathbf{324}$, 
  $\cdots$ 
  \\ \hline
$\mathrm{E}_6$
  & 
  $\mathbf{27}$, 
  $\mathbf{351}$, 
  $\mathbf{351'}$
  $\cdots$ 
  & 
  $\mathbf{78}$, 
  $\mathbf{650}$, 
  $\cdots$ 
  \\ \hline
$\mathrm{E}_7$
  & 
  & 
  $\mathbf{133}$ 
  $\cdots$
  \\ \hline
$\mathrm{E_8}$
  &
  & 
  $\mathbf{248}$, 
  $\cdots$ 
  \\ \hline
\end{tabular}
\end{scriptsize}
\end{center}

\end{table}

%
%


%
%
\section{Results}

%
%

Now we are ready to investigate the representations for fermions and scalars
in four dimensions.
We first note that we need a $R_{2}$ singlet in $\mathrm{SO}(6)$ vector to
obtain the Higgs candidate $h_{g}$ (\textit{cf}. Eq.(\ref{eq:dec_adjG_direct})
and the discussion below).
We can thus exclude the candidates (i) and (iii) of $S/R$ in
Table~\ref{table_6dim_coset_spaces} (\textit{cf}. Table
\ref{so6_vector_and_spinor}).
%
%
%
In Tables~\ref{SU5}--\ref{E6}, we list 
the possible candidates of $G_1$,
$G_2$, $(F_{1}, F_{2})$, and the corresponding representations of
four-dimensional scalars and fermions for each $H$, which is either 
$G_\mathrm{SM} \times \mathrm{U}(1)$, 
$\mathrm{SU}(5) \times \mathrm{U}(1)$, 
$\mathrm{SO}(10) \times \mathrm{U}(1)$,
or $\mathrm{E}_6 \times \mathrm{U}(1)$.
The representations of four dimensional fermions are classified into A, 
B, and C.
The representations of class A are the \textit{standard representations};
$\bar{\mathbf{5}}$ and $\mathbf{10}$ for $\mathrm{SU}(5)$, %
$\mathbf{16}$ for $\mathrm{SO}(10)$, and %
$\mathbf{27}$ for $\mathrm{E}_{6}$, %
which lead to the SM fermions after GUT breaking. 
The representations of class B lead to both of the SM fermions and non-SM fermions
after GUT breaking.
The representations of class C lead only to non-SM fermions after GUT breaking.

\subsection{$H=G_\mathrm{SM} \times \mathrm{U}(1)$}
We investigate all combinations of $S/R$, $G_1$ and $G_2$ in
Table~\ref{G_candidates_12}--\ref{G_candidates_4-cd} which provide $H =
G_{\mathrm{SM}} \times \mathrm{U}(1)$ in four 
dimensions.
We obtain a representation which is identified as the SM Higgs-doublet in four
dimensions from the following cases.
\begin{enumerate}
\item 
$R$ embedded as (ii)-(b), $G_1=\mathrm{SU}(6)$ and $G_2=\mathrm{SU}(2)$.

\item 
$R$ embedded as (ii)-(b), $G_1=\mathrm{SO}(11)$ and $G_2=\mathrm{SU}(2)$.

\item
$R$ embedded as (iv)-(c), $G_1=\mathrm{SU}(6)$ and $G_2=\mathrm{G_2}$.

\item
$R$ embedded as (iv)-(c), $G_1=\mathrm{SU}(6)$ and $G_2=\mathrm{Sp}(4)$.

\item
$R$ embedded as (iv)-(c), $G_1=\mathrm{SO}(11)$ and $G_2=\mathrm{G_2}$.

\item
$R$ embedded as (iv)-(c), $G_1=\mathrm{SO}(11)$ and $G_2=\mathrm{Sp}(4)$.

\item
$R$ embedded as (iv)-(d), $G_1=\mathrm{Sp}(12)$ and $G_2=\mathrm{SU}(2)$.

\item
$R$ embedded as (iv)-(d), $G_1=\mathrm{E_6}$ and $G_2=\mathrm{SU}(2)$.
\end{enumerate}
Any of these cases does not reproduce a whole generation of the SM fermions.
Therefore we cannot obtain the SM in four 
dimensions. 
The difficulty in obtaining the SM is ultimately due to the smallness of the
dimension of $\mathrm{SO}(6)$ spinor representation.

\subsection{$H=\mathrm{SU}(5) \times \mathrm{U}(1)$}
We investigate the case of $H=\mathrm{SU}(5) \times \mathrm{U}(1)$ and summarize
the result in Table~\ref{SU5}.
We obtain the representation $\mathbf{5}$ which corresponds to the Higgs scalar
in the following cases.
\begin{enumerate}
\item \label{SU5__2b_SU6_SU2}
$R$ embedded as (ii)-(b), $G_1=\mathrm{SU}(6)$ and $G_2=\mathrm{SU}(2)$.

\item \label{SU5__2b_SO11_SU2}
$R$ embedded as (ii)-(b), $G_1=\mathrm{SO}(11)$ and $G_2=\mathrm{SU}(2)$.

\item \label{SU5__2b_SU6_Sp4}
$R$ embedded as (iv)-(c), $G_1=\mathrm{SU}(6)$ and $G_2=\mathrm{Sp}(4)$.

\item \label{SU5__2b_SO11_Sp4}
$R$ embedded as (iv)-(c), $G_1=\mathrm{SO}(11)$ and $G_2=\mathrm{Sp}(4)$.

\item \label{SU5__2b_E6_SU2}
$R$ embedded as (iv)-(d), $G_1=\mathrm{E_6}$ and $G_2=\mathrm{SU}(2)$.
\end{enumerate}
As for the fermions, we see that the \textit{standard representations} of $\mathrm{SU}(5)$ GUT are
not obtained at all for the cases \ref{SU5__2b_SU6_Sp4}, \ref{SU5__2b_SO11_Sp4},
and \ref{SU5__2b_E6_SU2}, while they are obtained by combining two
representations of $F$ in the cases~\ref{SU5__2b_SU6_SU2} and
\ref{SU5__2b_SO11_SU2}.
For the example of case \ref{SU5__2b_SU6_SU2}, we can choose $(\mathbf{70},
\mathbf{2})$ and $(\mathbf{280}, \mathbf{1})$ to obtain all the \textit{standard
  representations}, $\mathbf{\bar{5}}$ and $\mathbf{10}$, in four 
dimensions,
along with the extra fermions of class B and C.

\subsection{$H=\mathrm{SO}(10) \times \mathrm{U}(1)$}
We investigate all the combinations of $S/R$, $G_1$, and $G_2$ for $H =
\mathrm{SO}(10) \times \mathrm{U}(1)$.
We obtain the representation $\mathbf{10}$ which corresponds to the Higgs scalar
in the following cases.
\begin{enumerate}
\item \label{SO10__2b_SO12_SU2}
$R$ embedded as (ii)-(b), $G_1=\mathrm{SO}(12)$ and $G_2=\mathrm{SU}(2)$.

\item \label{SO10__2b_E6_SU2}
$R$ embedded as (ii)-(b), $G_1=\mathrm{E_6}$ and $G_2=\mathrm{SU}(2)$.

\item \label{SO10__4b_SO14_SU2}
$R$ embedded as (iv)-(b), $G_1=\mathrm{SO}(14)$ and $G_2=\mathrm{SU}(2)$.

\item \label{SO10__4b_SO14_U1}
$R$ embedded as (iv)-(b), $G_1=\mathrm{SO}(14)$ and $G_2=\mathrm{U}(1)$.

\item \label{SO10__4b_SO15_SU2}
$R$ embedded as (iv)-(b), $G_1=\mathrm{SO}(15)$ and $G_2=\mathrm{SU}(2)$.

\item \label{SO10__4b_SO15_U1}
$R$ embedded as (iv)-(b), $G_1=\mathrm{SO}(15)$ and $G_2=\mathrm{U}(1)$.

\item
$R$ embedded as (iv)-(c), $G_1=\mathrm{SO}(12)$ and $G_2=\mathrm{G_2}$.

\item \label{SO10__4c_SO12_Sp4}
$R$ embedded as (iv)-(c), $G_1=\mathrm{SO}(12)$ and $G_2=\mathrm{Sp}(4)$.

\item
$R$ embedded as (iv)-(c), $G_1=\mathrm{E_6}$ and $G_2=\mathrm{SU}(2)$

\item
$R$ embedded as (iv)-(c), $G_1=\mathrm{E_6}$ and $G_2=\mathrm{G_2}$.

\item \label{SO10__4d_SO15_SU2}
$R$ embedded as (iv)-(d), $G_1=\mathrm{SO}(15)$ and $G_2=\mathrm{SU}(2)$.

\item \label{SO10__4d_E7_SU2}
$R$ embedded as (iv)-(d), $G_1=\mathrm{E_7}$ and $G_2=\mathrm{SU}(2)$.
\end{enumerate}
We further obtain the \textit{standard representations} of the fermions which
lead to all the SM fermions of one generation in the
cases~\ref{SO10__2b_SO12_SU2} -- \ref{SO10__4b_SO15_U1},
\ref{SO10__4c_SO12_Sp4}, \ref{SO10__4d_SO15_SU2} and \ref{SO10__4d_E7_SU2} (see
Table~\ref{SO10}).

The case~\ref{SO10__4b_SO14_U1} with $F = \mathbf{832} (1)$ is intriguing since
we obtain two $\mathbf{16}$s and two $\mathbf{144}$s, each of which 
leads to a
complete set of the SM fermions of one generation.
We thus obtain four generations of fermions which can accommodate the known
three generations.
Furthermore these representations can form three distinct types of Yukawa
coupling: %
$\mathbf{16} \times \mathbf{16} \times \mathbf{10}$, %
$\mathbf{144} \times \mathbf{16} \times \mathbf{10}$, and %
$\mathbf{144} \times \mathbf{144} \times \mathbf{10}$.
These couplings may explain the origin of the Yukawa couplings distinguishing the
the generations and the mixing among them.

\subsection{$H=\mathrm{E}_6 \times \mathrm{U}(1)$}
The results for $H=\mathrm{E}_6 \times \mathrm{U}(1)$ are listed in
Table~\ref{E6}.
We obtain representation $\mathbf{27}$ which corresponds to 
the Higgs scalar in
the following cases.
\begin{enumerate}
\item \label{E6__2b_E7_SU2}
$R$ embedded as (ii)-(b), $G_1=\mathrm{E_7}$ and $G_2=\mathrm{SU}(2)$.

\item
$R$ embedded as (iv)-(c), $G_1=\mathrm{E_7}$ and $G_2=\mathrm{G_2}$.

\item \label{E6__4d_E8_SU2}
$R$ embedded as (iv)-(d), $G_1=\mathrm{E_8}$ and $G_2=\mathrm{SU}(2)$.
\end{enumerate}
The \textit{standard representations} of fermion $\mathbf{27}$, which provide
all the SM fermions of one generation, 
are obtained in 
cases~\ref{E6__2b_E7_SU2}
and \ref{E6__4d_E8_SU2}.

Case~\ref{E6__2b_E7_SU2} with $F = (\mathbf{133}, \mathbf{1})$ is
interesting since the structure of the SM with three 
generations may be
explained.
The Yukawa coupling of this model needs to be in the form
\begin{math}
  \overline{\mathbf{27}}(-2) \times \mathbf{27}(2) \times \mathbf{78}(0).
\end{math}
The fermion representation $\mathbf{27} + \mathbf{78}$ of $\mathrm{E}_{6}$
contains three generations of $\bar{\mathbf{5}} + \mathbf{10}$ in terms of its
$\mathrm{SU}(5)$ subgroup, giving the origin of the known three generations.
Indeed, this fermion content is analyzed in, for example, nonlinear sigma models
giving a family unification~\cite{Buchmuller:1983iu} based on a broken
$\mathrm{E}_{7}$ symmetry~\cite{Kugo:1983ai}, under which a reproduction of the
observed mixing structure among the three generations of fermions has been
attempted~\cite{Sato:1997hv}.
  %

\renewcommand{\arraystretch}{1.3}

\begin{table}[p]
  \caption{ %
    The models for $H$=SU(5) $\times$ U(1) which include
    the SM Higgs-doublet and one generation of the SM fermions in four dimensions.
    The fermions in four dimensions are classified into A, 
B, and C.
    The fermion-As contain only the SM fermions;
    fermion-Bs contain both the SM fermions and extra fermions;
    fermion-Cs contain only extra fermions.
  } 
\label{SU5}
\begin{footnotesize}

\begin{tabular}{|c|c|c||c|c|c|c|} \hline
\multicolumn{7}{c}{
$S/R$ = Sp(4)/[SU(2) $\times$ U(1)], \hspace{2mm}
$G_1$ $\supset$ SU(5) $\times$ U(1), \hspace{2mm}
$G_2$ $\supset$ SU(2)} \\ \hline \hline
$G_1$ 	&$G_2$	&($F_1$, $F_2$)	&Scalars						&Fermions-A & B & C	\\ \hline \hline
SU(6)	&SU(2)	&(56, 2)			&5(6), $\bar{5}$(-6)			& &15(-3) & 35(-3) \\ \cline{3-7}
		&		&(70, 2)			&5(6), $\bar{5}$(-6)			& 10(-3)& 15(-3), 40(-3) & \\ \cline{3-7}
		&		&(280, 1)			&5(6), $\bar{5}$(-6)			&$\bar{5}$(-6) & $\overline{70}$(-6)  & 24(0), 45(-6), 126(0) \\
		&		&					&								& & &   24(0), 126(0) \\ \cline{3-7}
		&		&(405, 1)			&5(6), $\bar{5}$(-6)     &$\bar{5}$(-6) & $\overline{70}$(-6) & 1(0), 24(0), 200(0) \\ \cline{5-7} 
		&		&					&					     & & &  5(6), 70(6), 1(0), 24(0), 200(0) \\ \cline{3-7}
		&		&(840, 1)			&5(6), $\bar{5}$(-6)			& & $\overline{45}(6)$  & 280'(6), 126(0), 224(0) \\
		&		&					&								& & & $\overline{105}(6)$, 126(0), 224(0) \\ \hline
SO(11)	&SU(2)	&(11, 1)			&5(2), $\bar{5}$(-2)		& $\bar{5}$(-2) & & 1(0) \\ \cline{3-7} 
		&		&(55, 1)			&5(2), $\bar{5}$(-2)		& $\bar{5}$(-2) & & 1(0), 24(0) \\ \cline{3-7}
		&		&(65, 1)			&5(2), $\bar{5}$(-2)		&$\bar{5}$(-2) & & 1(0), 24(0) \\ \cline{3-7}
		&		&(165, 1)			&5(2), $\bar{5}$(-2)		&$\bar{5}$(-2) & $\bar{45}$(-2) & 1(0), 24(0) \\ \cline{3-7}
		&		&(275, 1)			&5(2), $\bar{5}$(-2)		&$\bar{5}$(-2) & $\bar{70}$(-2) & 1(0), 24(0) \\ \cline{3-7}
		&		&(320, 2)			&5(2), $\bar{5}$(-2)		& 10(-1), 10(-1) & 15(-1), 40(-1) &  \\ \cline{3-7}
		&		&(330, 1)			&5(2), $\bar{5}$(-2)		&$\bar{5}$(-2) &   $\bar{45}$(-2)& 1(0), 24(0), 75(0) \\ \cline{3-7}
		&		&(429, 1)			&5(2), $\bar{5}$(-2)		& $\bar{5}$(-2), $\bar{5}$(-2) & $\bar{45}$(-2), $\bar{70}$(-2) & 
																		1(0), 24(0), 24(0) \\ \cline{3-7}
		&		&(462, 1)			&5(2), $\bar{5}$(-2)		&$\bar{5}$(-2) & $\bar{45}$(-2), $\bar{50}$(-2) & 1(0), 24(0),
                                                                      75(0) \\ \cline{3-7}
		&		&(935, 1)			&5(2), $\bar{5}$(-2)		&$\bar{5}$(-2) & $\bar{70}$(-2) & 1(0), 24(0), 200(0) \\ \hline
\end{tabular}
\end{footnotesize}
\end{table}


\begin{table}[p]
  \caption{ %
    The models for $H = \mathrm{SO}(10) \times \mathrm{U}(1)$ which
    include the SM Higgs and one generation of the SM fermions in
    four-dimensions.
    The fermions in four-dimensions are classified into A, 
B, and C  where
    fermion-As are $\mathbf{16}$ representation of $\mathrm{SO}(10)$;
    fermion-Bs contain  both the SM fermions and extra-fermions;
    fermion-Cs contain only extra-fermions.
    We can obtain two types of results for fermions from one combination of
    $(G_{1}, G_{2}, F)$ since we have a freedom
    to change the overall sign of $\mathrm{U}(1)$ charges which appear in the
    $R$-decomposition of $\mathrm{SO}(6)$ vector and spinor.
  }
\label{SO10}
\begin{footnotesize}
\begin{tabular}{|c|c|c||c|c|c|c|} \hline \hline
\multicolumn{7}{c}{$S/R$ = Sp(4)/[SU(2) $\times$ U(1)]. $G_1$ $\supset$ SO(10) $\times$ U(1). $G_2$ $\supset$ SU(2)} \\ \hline \hline
$G_1$ 	&$G_2$	&($F_1$, $F_2$)	&Scalars						&Fermions-A & B & C	\\ \hline 
SO(12)	&SU(2)			&(12, 1)			&10(2), 10(-2)					& & 10(0),  &1(2) \\ \cline{5-7}
		&				&					& 								& & 10(0),  &1(-2) \\ \cline{3-7}		
		&				&(66, 1)			&10(2), 10(-2)					& &10(2), 45(0) & 1(0) \\ \cline{5-7}  
		&				&					&								& &10(-2), 45(0) & 1(0) \\ \cline{3-7}		
		&				&(77, 1)			&10(2), 10(-2)					& &10(2), 54(0) & 1(0) \\ \cline{5-7} 
		&				&					&								& &10(-2), 54(0) &1(0) \\ \cline{3-7}		
		&				&(220, 1)			&10(2), 10(-2)					& & 45(2), 10(0), 120(0) & \\ \cline{5-7}  
		&				&					&								& &45(-2), 10(0), 120(0) & \\ \cline{3-7}		
		&				&(352, 1)			&10(2), 10(-2)					& & 54(2), 10(0), 210'(0) & \\ \cline{5-7} 
		&				&					&								& &54(-2), 1(-2), 10(0), 210'(0) & \\ \cline{3-7}		
		&				&(462, 1)			&10(2), 10(-2)					& & 126(2), 210(0) & \\ \cline{5-7} 
		&				&					&								& &	$\bar{126}$(-2), 210(0) & \\ \cline{3-7}		
		&				&(495, 1)			&10(2), 10(-2)					& & 120(2), 45(0), 210(0) & \\ \cline{5-7}
		&				&					&								& &	120(-2), 45(0), 210(0) & \\ \cline{3-7}
		&				&(560, 1)			&10(2), 10(-2)					& & 54(2), 45(2), 10(0), 10(0), 320(0) & 1(2) \\ \cline{5-7}
		&				&					&							& &54(-2), 45(-2), 10(0), 10(0), 320(0)&1(-2) \\ \cline{3-7}
		&				&(792, 1)			&10(2), 10(-2)					& & 210(2), 120(0), 126(0), $\bar{126}$(0)& \\ \cline{5-7}
		&				&					&								& &210(-2), 120(0), 126(0), $\bar{126}$(0)& \\ \hline
E$_6$	&SU(2)			&(78, 1)			&16(-3), $\bar{16}$(3)			&16(-3)&  45(0) &1(0) \\ \cline{5-7}
		&				&					&								&	&  45(0) & 1(0) $\bar{16}$(3)\\ \cline{3-7}
		&				&(650, 1)			&16(-3), $\bar{16}$(3)	    &16(3)& $\bar{144}$(3), 45(0), 54(0), 210(0)& 1(0) \\ \cline{5-7}
		&				&					&							& &  144(-3),45(0), 54(0), 210(0)&1(0) $\bar{16}$(-3) \\ 
\hline \hline
\multicolumn{7}{l}{
$S/R$ = Sp(4) $\times$ SU(2)/[SU(2) $\times$ SU(2)] $\times$ U(1), \hspace{2mm}
$G_1$ $\supset$ SO(10) $\times$ SU(2) $\times$ SU(2), \hspace{2mm}
$G_2$ $\supset$ U(1)} \\ \hline \hline
$G_1$ 	&$G_2$	&($F_1$, $F_2$)	&Scalars						&Fermions-A & B & C 	\\ \hline 
SO(14)	&SU(2)			&(64, 2)	&10(0)				&16(1),16(1)& & $\overline{16}$(-1),$\overline{16}$(-1)  \\ \cline{2-7}		
		&U(1)			&64(1)		&10(0)			    &16(1),16(-1) & &  \\ \cline{3-7}		
		&				&832(1)		&10(0)				&16(1),16(-1)& 144(1) {144}(-1)&   \\ \hline		
SO(15)	&SU(2)			&(128, 2)	&10(0), 1(0)				&16(1), 16(-1)& & $\overline{16}$(1), $\overline{16}$(-1)  \\ \cline{2-7}
		&U(1)			&128(1)		&10(0), 1(0)					&16(1)& &$\overline{16}$(1) \\ \hline
\multicolumn{7}{c}{$S/R$ = Sp(4) $\times$ SU(2)/[SU(2) $\times$ SU(2)] $\times$ U(1).　
$G_1$ $\supset$ SO(10) $\times$ SU(2) $\times$ U(1).　
$G_2$ $\supset$ SU(2)} \\ \hline \hline
$G_1$ 	&$G_2$	&($F_1$, $F_2$)	&Scalars						&Fermions-A & B & C 	\\ \hline 
SO(15) &SU(2)	&(128,1) &10(2),10(-2) &
16(1)& $\bar{16}$(1)&  \\ \hline		
$\mathrm{E}_{7}$	&SU(2)	&	(133,1)             &10(2),10(-2)                   &16(1)& & \\ \hline			
\end{tabular} 
\end{footnotesize}
\end{table}

\clearpage


\begin{table}[p]
  \caption{The models for $H = \mathrm{E}_{6}\times \mathrm{U}(1)$ 
    which include the SM Higgs and one generation of the SM fermions in
    four dimensions.
    The fermions in four dimensions are classified into A, B and C where
    fermion-As are $\mathbf{27}$ representation of $\mathrm{E}_{6}$;
    fermion-Bs contain both the SM fermions and extra fermions;
    fermion-Cs contain only extra-fermions.
    We can obtain two types of results for fermions from one combination of
    $(G_{1}, G_{2}, F)$ since we have a freedom to change the 
    overall sign of U(1) charges which appear in $R$-decomposition of
    $\mathrm{SO}(6)$ vector and spinor.
  }
\begin{tabular}{|c|c|c|c|c|c|c|} \hline \hline
\multicolumn{7}{c}{$S/R$ = Sp(4)/SU(2) $\times$ U(1), \hspace{2mm}
$G_1$ $\supset$ E$_6$ $\times$ U(1), \hspace{2mm}
$G_2$ $\supset$ SU(2)} \\ \hline \hline
$G_1$ 	&$G_2$	&($F_1$, $F_2$)	&Scalars						&Fermions-A &B &C	\\ \hline 
E$_7$   &SU(2)  &(133,1) & 27(2),$\bar{27}(-2)$ & 27(2) & 78(0) & 1(0)  \\ \cline{5-7} 
        &  		&        &                     &       & $\overline{27}$(-2), 78(0)  & 1(0)  \\ \hline \hline
\multicolumn{7}{c}{$S/R$ = Sp(4) $\times$ SU(2)/[SU(2) $\times$ SU(2)] $\times$ U(1), \hspace{2mm}
$G_1$ $\supset$ E$_6$ $\times$ SU(2) $\times$ U(1), \hspace{2mm}
$G_2$ $\supset$ SU(2)} \\ \hline \hline
$G_1$ 	&$G_2$	&($F_1$, $F_2$)	&Scalars						&Fermions-A & B & C	\\ \hline 
E$_8$	&SU(2)	&F(248, 1)			&27(-2), $\bar{27}$(2) 			&27(1) & & \\ \cline{5-7}
        &       &                   &                               &      & $\bar{27}$(-1) & \\ \hline        
\end{tabular}
\label{E6}
\end{table}

\section{Summary and Discussions}
\label{sec:Summaries-Discussions}
We studied 
the ten-dimensional gauge theories whose extra six-dimensional spacetime
is a coset space of Lie groups.
We focused on the case where the gauge group is a direct product of two simple
Lie groups, and searched for models which lead to phenomenologically promising
four-dimensional models after applying the coset space dimensional reduction.

We first limited the possible coset space $S/R$ to four types listed in (i) --
(iv) of Table~\ref{table_6dim_coset_spaces} by requiring that $R$ should be
factored as $R = R_{1} \times R_{2}$.
All 
of these four types have a $\mathrm{U}(1)$ factor in $R$, but this
$\mathrm{U}(1)$ can never be identified as the hypercharge symmetry of the SM.
We thus needed to introduce an extra $\mathrm{U}(1)$ in the four-dimensional
gauge group $H$, and searched for SM-like models or GUT-like ones.
The former is the case where %
\begin{math}
  H =
  \mathrm{SU}(3)
  \times \mathrm{SU}(2)
  \times \mathrm{U}(1)
  \times \mathrm{U}(1),
\end{math}
while the latter is where
$H = \mathrm{SU}(5) \times \mathrm{U}(1)$,
$H = \mathrm{SO}(10) \times \mathrm{U}(1)$,
and
$H = \mathrm{E}_{6} \times \mathrm{U}(1)$.
We also require that the induced four-dimensional model should include the
particle contents appropriate for the SM particles.
We then found the candidates of the gauge group $G = G_{1} \times G_{2}$ of the
ten-dimensional theory and the representations for fermions.

For each of the obtained candidates, we made the complete lists of
representations of the scalars and the fermions that constitute the
corresponding four-dimensional theory.
The results are summarized as follows.
\begin{enumerate}
\item No ten-dimensional model was found to induce the promising model with 
  \begin{math}
    H =
    \mathrm{SU}(3) \times \mathrm{SU}(2) \times
    \mathrm{U}(1) \times \mathrm{U}(1)
\end{math}
in the four-dimensional spacetime.
\item The models which induce a $\mathrm{SU}(5) \times \mathrm{U}(1)$ gauge
  theory in four-dimensional spacetime 
were found when $S/R = \mathrm{Sp}(4) /
  \mathrm{SU}(2) \times \mathrm{U}(1)$.
  Possible gauge group is either $\mathrm{SU}(6) \times \mathrm{SU}(2)$ or
  $\mathrm{SO}(11) \times \mathrm{SU}(2)$, and each case has several choices of
  the representation for the ten-dimensional fermions as listed in
  Table~\ref{SU5}.
  Many of fermion representations generate either $\bar{\mathbf{5}}$ or
  $\mathbf{10}$ of the $\mathrm{SU}(5)$ after the dimensional reduction.
  None of them, however, generates both from a single representation, and we
  thus need at least two fermion representations in the ten-dimensional model as
  well as in four-dimensional one.
\item The models which induce a $\mathrm{SO}(10) \times \mathrm{U}(1)$ gauge
  theory in four dimensions 
were found for the three possible 
choices
  of $S/R$, and each choice allows a number of gauge groups as listed in
  Table~\ref{SO10}.
\item The models which induce a $\mathrm{E}_{6} \times \mathrm{U}(1)$ gauge
  theory in four dimensions 
were found when %
  \begin{math}
    S/R = \mathrm{Sp}(4)/\mathrm{SU}(2) \times \mathrm{U}(1), \,
    G = \mathrm{E}_{7} \times \mathrm{SU}(2)
  \end{math}
  and
  \begin{math}
    S/R = \mathrm{Sp}(4) \times \mathrm{SU}(2)/
          [\mathrm{SU}(2) \times \mathrm{SU}(2)] \times \mathrm{U}(1), \,
    G = \mathrm{E}_{8} \times \mathrm{SU}(2),
  \end{math}
  as listed in Table~\ref{E6}.
\end{enumerate}
The fermion representations in four-dimensional theories obtained from the
candidate models mentioned above are not limited to the standard ones,
\textit{i.e.}  $\bar{\mathbf{5}}$ and $\mathbf{10}$ for $H = \mathrm{SU}(5)
\times \mathrm{U}(1)$, $\mathbf{16}$ for $H = \mathrm{SO}(10) \times
\mathrm{U}(1)$, and $\mathbf{27}$ for $H = \mathrm{E}_{6} \times \mathrm{U}(1)$.
Some of these extra representations can accommodate the SM particles as well and
thus can take part in the further model buildings.
The following two models are found to be of particular interest.
\begin{enumerate}
\item %
  $H = \mathrm{SO}(10) \times \mathrm{U}(1)$,
  \begin{math}
    S/R = \mathrm{Sp}(4) \times \mathrm{SU}(2) /
    [\mathrm{SU}(2) \times \mathrm{SU}(2)] \times \mathrm{U}(1),
  \end{math}
  $G = \mathrm{SO}(14) \times \mathrm{U}(1)$, and $F = \mathbf{832} (1)$ (see
  Table~\ref{SO10}).
  In this case, the fermions in four-dimensional theory 
include two
  $\mathbf{16}$s and two $\mathbf{144}$s.
  Since both can include a complete set of the SM fermions of a generation, this
  case has four generations of fermions and thus can accommodate the known three
  generations.
  Besides, this case allows three distinct types of Yukawa coupling:
  $\mathbf{16} \times \mathbf{16} \times \mathbf{10}$,
  $\mathbf{144} \times \mathbf{16} \times \mathbf{10}$, and
  $\mathbf{144} \times \mathbf{144} \times \mathbf{10}$.
  These three types of couplings can admit the different Yukawa couplings,
  giving rise to the distinction of generations.
  Hence this model may possibly introduce the mixing among the generations.
  \item %
  $H = \mathrm{E}_{6} \times \mathrm{U}(1)$,
  $S/R = \mathrm{Sp}(4) / \mathrm{SU}(2)$,
  $G = \mathrm{E}_{7} \times \mathrm{SU}(2)$,
  and $F = (\mathbf{133}, \mathbf{1})$
  (see Table~\ref{E6}).
  The Yukawa coupling of this model is necessarily of the form
  \begin{math}
    \overline{\mathbf{27}}(-2) \times \mathbf{27}(2) \times \mathbf{78}(0).
  \end{math}
  The fermion representation $\mathbf{27} + \mathbf{78}$ of $\mathrm{E}_{6}$
  contains three generations of $\bar{\mathbf{5}} + \mathbf{10}$ in terms of its
  $\mathrm{SU}(5)$ subgroup, giving the origin of the known three generations.
  Indeed, this fermion content is analyzed in, for example, nonlinear sigma
  models giving a family unification~\cite{Buchmuller:1983iu} based on a broken
  $\mathrm{E}_{7}$ symmetry~\cite{Kugo:1983ai}, under which a reproduction of
  the observed mixing structure among the three generations of fermions has been
  attempted~\cite{Sato:1997hv}.
\end{enumerate}
We leave further analysis for the future study as well as building
phenomenological models based on the models mentioned above.
%


\acknowledgments

The work of T.~J. was financially supported by the Sasakawa Scientific Research 
Grant from the Japan Science Society. 
The work of T.~N. was supported in part by the Grant-in-Aid for the Ministry
of Education, Culture, Sports, Science, and Technology, Government of
Japan (No. 19010485). 
The work of J.~S. was supported in part by the Grant-in-Aid for the Ministry
of Education, Culture, Sports, Science, and Technology, Government of
Japan (No. 20025001, 20039001, and 20540251).
The work of T.~S. was supported in part by MEC and FEDER (EC), Grants No. FPA2005-01678 and
the Generalitat Valenciana for support under the grants GVPRE/2008/003, PROMETEO/2008/004
and GV05/267.


\end{document}